\documentclass[journal,a4paper]{IEEEtran}

\IEEEoverridecommandlockouts
\usepackage[x11names,table,xcdraw]{xcolor}
\usepackage{amsmath,amssymb,amsfonts}
\usepackage{algorithmic}
\usepackage{graphicx}
\usepackage{subfig}
\usepackage{textcomp}
\usepackage{enumitem}
\usepackage{romanbar}
\usepackage{tikz}
\usepackage{mathdots}
\usepackage{yhmath}
\usepackage{cancel}
\usepackage{array}
\usepackage{multirow}
\usepackage{gensymb}
\usepackage{tabularx}
\usepackage{booktabs}
\usepackage{verbatim}
\usepackage{balance}
\usepackage{adjustbox}
\usepackage[linesnumbered,ruled,vlined]{algorithm2e}
\algsetup{linenosize=\small}
\usepackage{multicol}

\SetCommentSty{mycommfont}
\usetikzlibrary{fadings}

\def\BibTeX{{\rm B\kern-.05em{\sc i\kern-.025em b}\kern-.08em
    T\kern-.1667em\lower.7ex\hbox{E}\kern-.125emX}}

\usepackage{mathtools}  
\usepackage{makecell}   

\begin{document}

\title{Distributed Artificial Intelligence Solution for D2D Communication in 5G Networks
\thanks{This research is part of a project that has received funding from the European Union's Horizon 2020 research and innovation programme under grant agreement Nº739578 and the government of the Republic of Cyprus through the Directorate General for European Programmes, Coordination and Development.}
}

\author{Iacovos Ioannou, Vasos Vassiliou, Christophoros Christophorou, and Andreas Pitsillides\\
Department of Computer Science, University of Cyprus and \\
RISE Center of Excellence on Interactive Media, Smart Systems and Emerging Technologies\\
Nicosia, Cyprus \\
}

\maketitle
\bibliography{conference_ieee}

\begin{abstract}
Device to Device (D2D) Communication is one of the technology components of the evolving 5G architecture, as it promises improvements in energy efficiency, spectral efficiency, overall system capacity, and higher data rates.  The above noted improvements in network performance spearheaded a vast amount of research in D2D, which have identified significant challenges that need to be addressed before realizing their full potential in emerging 5G Networks. Towards this end, this paper proposes the use of a distributed intelligent approach to control the generation of D2D networks. More precisely, the proposed approach uses Belief-Desire-Intention (BDI) intelligent agents with extended capabilities (BDIx) to manage each D2D node independently and autonomously, without the help of the Base Station. This paper proposes the DAIS algorithm  for the decision of transmission mode in D2D, which maximizes the data rate and minimizes the power consumption, while taking into consideration the computational load. Simulations show the applicability of BDI agents in solving D2D challenges.

\end{abstract}


\section{Introduction}
\label{intro}
Device to Device (D2D) Communication is expected to be a core part of the forthcoming 5G mobile communication networks. D2D can operate both in the licensed and unlicensed spectrum and is generally transparent to the cellular network as it allows adjacent user equipment (UE) to bypass the base station (BS) and establish direct links between them, to either share their connection and act as relay stations, or directly communicate and exchange information.
D2D can be used to implement many of the 5G requirements, because it can support high bit rates and minimize the delay between D2D UEs. The gains of D2D communications in spectral efficiency,  resource reallocation, and reduction of interference \cite{b1,b2} can potentially improve throughput, energy efficiency, delay, and fairness \cite{b3,b4}. In addition, due to the shorter communication distance, D2D can offer lower power consumption for the communicating D2D devices.
D2D can enable mobile traffic offloading, so overall one can anticipate that the non-D2D UEs can also benefit from the mobile traffic offloading because they will, as a result, have access to more bandwidth for the communication between them (non-D2D UEs) and the BS, as well as less interference \cite{b3,b4}.
However, in order to fully realize D2D, several challenges need to be resolved, including device discovery, mode selection, interference management, power control, security, radio resource allocation, cell densification \& offloading, Quality of Service (QoS) \& path selection, use of mmWave communication, non-cooperative users, and handover management \cite{b26,b36,b37}.

This work investigates the idea that the D2D communication is not a global problem that must be solved centrally, but it is an optimization problem that should be solved in a distributed fashion with the use of artificial intelligence. To address that, the paper proposes that the control is handled by the UEs, locally, in order to form communication links in shorter time \cite{b5,b6,b7,b8,b9,b10,b11,b12,b13}. We consider that the use of distributed artificial intelligence (AI) control is the most suitable in the challenging and dynamic environment of D2D communication. To the best of our knowledge, there are no solutions in the literature that jointly satisfy all of the D2D requirements in one approach. We chose intelligent agents because of their ability to concurrently solve multiple complex problems, as it was shown in \cite{b40}.

In this paper we are making the following contributions:
\begin{enumerate}[label={(\alph*)}]
	\item we propose a solution using Belief-Desire-Intention (BDI) software agents with extended capabilities (BDIx), to collectively satisfy the challenges identified for D2D communication,
	\item we provide a proof-of-concept algorithm that encompasses the use of intelligent agents for selecting the D2D transmission mode, while ensuring a high spectral efficiency and low computational load,
    \item we propose the use of a new parameter called Weighted Data Rate (WDR) for the decision of D2D transmission mode, and
	\item we evaluate the proposed solution under varying scenarios and provide insights into its operation.
\end{enumerate}

The rest of the paper is structured as follows. Section II provides background information on D2D communications and intelligent agents. Section III discusses related work in AI techniques for communications and D2D. Section IV presents the proposed solution of distributed control in D2D through BDIx agents and describes the DAIS algorithm. Section V discusses the evaluation of the proposed approach, and lastly, Section VI contains our conclusions and ideas for future work.\\

\section{Background on D2D and BDI Agents}

\subsection{Background on D2D}\label{CD2D}

\subsubsection{Control of D2D Communication}
We can categorize the solutions on D2D communication based on the type of control, as follows:

\begin{itemize}
\item Centralized: In centralized techniques the BS completely manages the UE nodes, even when they (UEs) are communicating directly. The controller manages all aspects if interference/connections/path etc., between cell and D2D UEs.
\item Distributed: In a distributed scheme, the procedure of D2D node management does not require a central entity, but it is performed autonomously by the UEs themselves. The distributed scheme decreases the control and computational overhead. This tactic is more suitable for large size D2D networks. In such a system, all control processes are run in parallel and start at the same time.
\item Semi Distributed: In spite of the fact that both centralized and distributed schemes have their strong points, tradeoffs can be accomplished between them. Such D2D  management schemes are referred to as semi-distributed" or "hybrid".
\end{itemize}

\subsubsection{Transmission Mode in D2D Communication}\label{TMode}
\begin{figure}[h]
	\centering
	\includegraphics[width=0.9\linewidth]{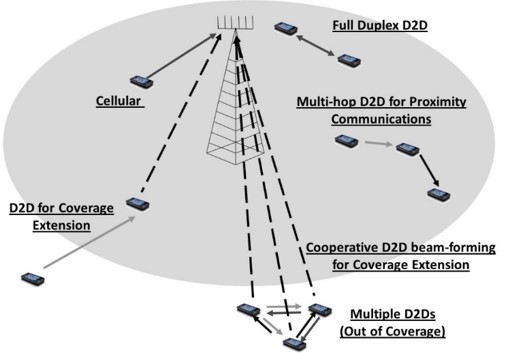}
	\caption[Transmission Mode in D2D Communication]{Transmission in D2D Communication}
	\label{fig:D2Dmode}
\end{figure}

There exist different modes for D2D communication, based on how UEs interact with the BS and other D2D nodes (see Fig. \ref{fig:D2Dmode}).
\begin{itemize}
	\item D2D Direct: Two UEs connect to each other by using licensed or unlicensed spectrum. The two D2D UEs only communicate with each other (also called Full-Duplex D2D).

	\item D2D Single-hop Relay: Sharing of bandwidth between a UE and other UEs. In D2D Single-hop Relay mode one of the D2D UEs is connected to a BS or access point (AP) and provides access to another D2D UE. \cite{b31}.

	\item D2D Multi-hop Relay: The single-hop mode is extended by enabling the connection of more D2D UEs in chain. Both backhaul and D2D transmissions are performed in an uplink with other D2D relay node (as a bridge) and they are subject to the control of the other D2D relay node \cite{b32}.		
	
	\item D2D Cluster: D2D cluster is a group of UEs connected to a D2D relay node acting as a Cluster Head (CH). The D2D relay node acts as an intermediate router to the network though an access point or BS. Clustering is suitable in high user densities \cite{b33,b34,b35}.
\end{itemize}

\subsection{Research Challenges in D2D}\label{RC}
In order for D2D to mature and shape the D2D communication for the upcoming 5G and beyond wireless communication networks, some technical issues must be resolved \cite{b36,b37}. Each of these challenges is further elaborated below.

\subsubsection{Device Discovery}
In order for two devices (i.e., UEs) to directly communicate with one another, they must first perform a device discovery process to identify that they are close to each other and in range for D2D communication \cite{b2,b20}.

\subsubsection{Mode Selection}
When a pair of D2D candidates identify each other for possible future communication, mode selection is performed. Mode selection implies that a decision is made whether the D2D candidates will communicate directly or via the conventional cellular network \cite{b20}. The communication mode selection should be carefully chosen in order not to impact on the interference in the network. This communication mode decision is categorized in the following way:

\begin{enumerate}[label={(\alph*)}]
\item Inband D2D communication:
\begin{itemize}
\item \emph{Reuse/Underlay:} D2D communication shares the same resources with existing Cellular UEs. This mode can achieve high spectral efficiency; however, it may cause interference to other Cellular and D2D UEs using the cellular resources.
\item\emph{ Dedicated/Overlay:} The cellular network has abundant channel resources so that the D2D UEs can use dedicated resources that are orthogonal to cellular UEs.
\item \emph{Cellular:} The two UEs will communicate with each other via the cellular network as traditional cellular UEs.
\end{itemize}

\item Outband D2D Communication:
\begin{itemize}
\item \emph{Controlled:} In the controlled mode the device has two interfaces. On the first interface it uses unlicensed spectrum to share with its peers. On the second interface it uses licensed spectrum to connect to the mobile network.
\item \emph{Autonomous:} In autonomous mode, the device can only use and communicate with other devices under the unlicensed spectrum, without accessing BS.
\end{itemize}
\end{enumerate}

\subsubsection{Interference Management}
The communication mode selection has a direct impact on the interference in the network. For example when the Reuse/Underlay resource-sharing mode is selected, high spectral efficiency can be achieved. However, since many D2D and cellular users will use the same portion of spectrum, interference may become a problem. Therefore, interference management must be used \cite{b20}.

\subsubsection{Power Control}
Although high transmission power can provide wider coverage and better signal quality during D2D communication, it can, at the same time, drain the battery of D2D UEs and cause interference to the network. Thus, proper power control during D2D communication is vital for controlling the transmission power levels of D2D UEs so as to deal with the interference generated by the D2D UEs and improve spectral efficiency, system capacity, coverage, and reduce energy consumption \cite{b21,b22,b23}.

\subsubsection{Security Concerns}
In D2D communications, the routing of users’ data is done through other users’ devices. This makes the D2D communication network vulnerable to many security risks and malicious attacks that could breach the data privacy and confidentiality. Thus, providing efficient security is a major issue in order to facilitate D2D communication in cellular networks \cite{b24,b25,b12}.

\subsubsection{Radio Resource Allocation}
Radio resource allocation mainly addresses the issues of how to assign the frequency resources to a group of D2D pairs, or all the D2D pairs, targeting an optimal use of the radio resources focusing also on the interference control and management between D2D and cellular links and the efficient reuse the radio resources whenever the interference is small \cite{b20}.

\subsubsection{Cell Densification and Offloading}
Providing high system capacity and high per-user data rates – requirements for the creation of a 5G network – will require a densification of the radio access network or the deployment of additional network nodes. In general, the need of network densification \cite{b26} for performance enhancement dictates the deployment of small coverage cells \cite{b20}.

\subsubsection{QoS / Path Selection (Routing)}
During D2D communication it is essential to ensure that the QoS requirements of the communication links are satisfied. To achieve this a major issue to handle is the selection of the optimum routing path, otherwise excess resources/power/link usage (bandwidth) will be wasted \cite{b22,b27,b28}.

\subsubsection{D2D in mmWave communication}
Communication using the mmWave band has recently received significant attention for 5G cellular networks and D2D communication, as it operates at a higher frequency band (30-300 GHZ) and allows  a significant increase in data rates (multi-Gbps) and network capacity \cite{b29}.

\subsubsection{Handover of D2D device}
In order to keep the communication between two D2D devices when these are moving away from each other, handover should be performed. More specifically, when a D2D device is moving away from the access point (e.g., a D2D Relay or a D2D Cluster Head) it is assigned to, then the problem of handing it over to another access point (e.g., another D2D Relay or D2D Cluster Head) with a shared medium should be dealt with \cite{b22}.

\subsubsection{Non-cooperative Users}
An issue to consider for D2D data delivery is that the data delivery in non-cooperative D2D communication may be unfair or compromised. In the real world, some rational nodes may have strategic interactions and may act selfishly for various reasons (such as resource limitations, the lack of interest in data, or social preferences) or even malicious nodes that they may use the data relay to attack anonymously \cite{b30}.

\subsection{Background on Intelligent Agents and Belief-Desire-Intention Agents}
\subsubsection{Intelligent Agents} An intelligent agent (IA) is an autonomous unit, which observes an environment using sensors and acts upon it using actuators, coordinating its activity in the direction of achieving goals (i.e. it is "rational", as defined in economics) \cite{b15}. Agent theory is concerned with the use of mathematical formalisms for representing reasoning and the properties of agents.  Software agents are characterized as computer software that display flexible autonomous behavior, which infers that these systems are capable of independent, autonomous action in order to satisfy their design objectives. Agents are utilized in a lot of applications. For instance, autonomous programs used for operator assistance or data mining (in some cases referred as bots) are also called "intelligent agents".

\subsubsection{Belief-Desire-Intention Agents}
\begin{figure}
	\centering
	\includegraphics[width=0.9\linewidth]{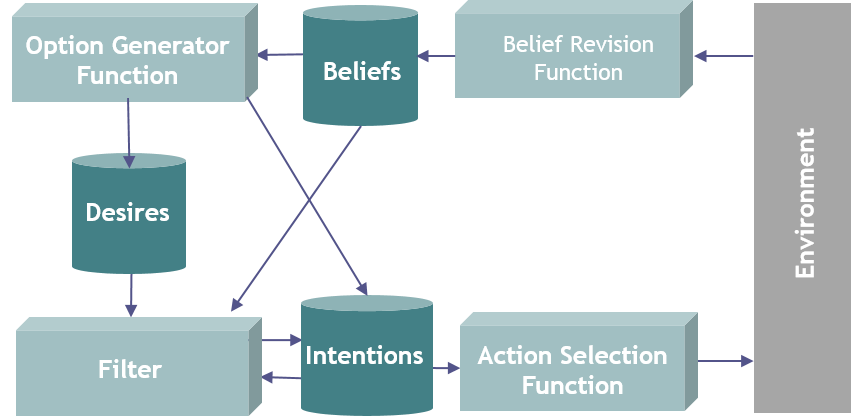}
	\caption[BDI Agent Architecture]{BDI Agent Architecture}
	\label{fig:AgentArch}
\end{figure}

This work makes use of Belief-Desire-Intention (BDI) software agents, which are agents with three key mental structures (see Fig. \ref{fig:AgentArch}): informative states of mind around the world (beliefs or convictions), motivational approaches on what to do (desires or wants) and planned responsibilities to take action (intentions or expectations). The BDI model fundamentally relies on two principle forms: thought and mean send thinking. With the thought processes the agent produces its goals on the premise of its convictions and desires, while mean send thinking comprises of a succession of activities to execute, as an endeavor to satisfy desires \cite{b16}.

Unique features of BDI agents \cite{b17}:
\begin{enumerate}[label={(\alph*)}]
	\item \textbf{Beliefs:} Beliefs correspond to the informational state of the agent. Beliefs can also include inference rules, allowing advance chaining to guide to new beliefs.
	\item \textbf{Desires:} Desires correspond to the motivational state of the agent. They characterize objectives or situations that the agent would like to fulfil or bring about.
	\item \textbf{Intentions:} Intentions correspond to the deliberative state of the agent. This is what the agent has chosen to perform. Intentions are desires to which the agent has, to some extent, committed.
\end{enumerate}

A BDI agent decides its actions based on beliefs, which either contribute to the achievement of its goals, or react to its received (or perceived) events and messages. \cite{b18}.
BDI agents can also cooperate and form a multi-agent system. Multi-agent systems are systems composed of multiple interacting computing elements capable of autonomously deciding what actions they require to perform in order to satisfy their design objectives. In multi-agent systems, the entities are interacting with other agents, not only by exchanging information, but also by applying analogues of the type of social activity that people engage in every day, like cooperation, coordination, and negotiation \cite{b18}.
In multi-agent systems, there are two important issues to consider: (a) since agents are anticipated to be autonomous, it is usually expected that the synchronization and coordination structures in a multi-agent system are not hard-wired at design time, as they normally are in standard concurrent/distributed systems. In this manner, mechanisms are needed in order to allow agents to synchronize and coordinate their activities at runtime; and (b) the encounters that occur between computing elements in a multi-agent system are financial encounters, in the sense that they are encounters between self-interested entities. In a classic distributed/concurrent system, all the computing elements are implicitly expected to share the common goal of making the overall system function correctly. In multi-agent systems, it is assumed instead, that agents are primarily concerned with their own welfare, although of course, they will be acting on behalf of some user/owner \cite{b18}.

In addition, we can say the BDI agents have foundations in the Algorithmic, Game-Theoretic, and Logical theories \cite{b18}. All the features discussed above make, in our opinion, BDI agents suitable for solving the challenges of D2D.

\section{Related Work}

\subsection{Related Work on AI Techniques for Communications}
There is a wealth of research on the use of artificial intelligence (AI) and machine learning (ML) techniques for communication and networking issues. In this section we include a few examples that deal with the use of multi-agent systems and BDI agents in general communication problems and at the end focus on AI approaches for D2D communication.

\subsection{Multi-agent Approaches for Wireless and Mobile Communications}
The authors in \cite{b40} address the problem of energy consumption and communication latency in wireless sensor networks (WSNs). More specifically, the authors propose a system with a single mobile agent (MA) travelling freely within the network and performing data collection. This behavior improves data delivery to the sink, and reduces energy consumption. The specific work utilizes deep neural network for learning, in which the input is the state of the wireless sensor network and the output is the optimal route path.  The route planning can be done with the usage of the locations of each node in the environment that act as input for the intelligent agent. The intelligent agent architecture selected is the actor network and a critic network. The information used comes from the whole network, but the decision is taken locally.

Another work that uses reinforcement learning is \cite{b38}, which deals with the problem of discovering low-level wireless communication schemes between two agents in a fully decentralized system. This is the type of problem considered in the DARPA Spectrum Collaboration Challenge (SC2). The proposed method employs policy gradients to learn an ideal bi-directional communication scheme. The approach places two agents against each other and show that the two actors are able to learn modulation schemes for communication while sharing only limited information and having no domain-specific knowledge about the task.

\subsection{BDI Agents for Wireless and Mobile Communications}
The authors in \cite{b39} utilize a multi-agent software design, dynamic analysis, and decentralized control in order to implement solutions for the complex distributed systems of WSNs. The paper's purpose is to create an autonomic system design for distributed nodes in a diverse and changing environment, that interact on top of a wireless communication channel for decentralized problem solving. Due to hardware limitations, the multi-agent system techniques and especially nodes (agents) are not deliberative (or strong) reasoning systems. The BDI agent model is used. The paper's authors implement two simple WSN test scenarios and show that BDI agents can perform basic WSN functions. In addition, the agents succeed in imitating some recognizable aspects of the system and show that the solution is adaptable to different scenarios. In the scenarios, five different agents are discussed. A problem of this approach is that a better method is needed for managing the size of the belief-base used in each agent, as this turns out to expand unboundedly in a case such as flooding.

Another class of wireless networks built dynamically in an ad hoc network manner with a large mobile user base is found in vehicular ad-hoc networks (VANETs). The work presented in \cite{b16} tackles the problem of routing in VANETs. Routing in VANETs is critical because of limitations such as unpredictable network topology, frequent disconnections, and varying network densities.  The authors in this paper proposed a multi-agent scheme-based routing scheme that comprises of static agent and mobile agents for vehicle-to-vehicle communication (V2V), where they address the challenge of how to route the data with short communication delay, overhead, and the complexity. The proposed algorithm has the following steps: i) establish a connectivity pattern between the vehicles; ii) create a set of beliefs; iii) develop the desires, and iv) execute the intentions.

\subsection{Artificial Intelligence Approaches for D2D}
In the last decade we have seen many approaches for solving the D2D challenges using AI and ML \cite{b50}.
The authors in \cite{b22} proposed EHSD -Exemplary Handover Scheme During D2D Communication- a framework describing a handover scheme that is based on software-defined radio (SDR) decentralization by using fuzzy logic. In \cite{b43}, the authors proposed a learning-based resource allocation approach for D2D communications with QoS and fairness considerations by using Q-Learning. In addition, in \cite{b44} the authors proposed a Hierarchical Extreme Learning Machine (H-ELM) Neural Network in order to manage the severe interference in D2D communications. Another paper, \cite{b45}, proposed a genetic algorithm (GA)-based scheme for Fair Joint Channel Allocation and Power Control for Underlaying D2D Multicast Communications. Also in \cite{b46}, the authors proposed an approach for power control in two-tier orthogonal frequency division multiple access (OFDMA) femtocell networks by using particle swarm optimization (PSO). Another intelligent technique is presented in \cite{b47}, where the authors used an ant-colony optimization (ACO)-based resource allocation scheme for solving the problem of swarm intelligence-based radio resource management for D2D-based V2V communication.

In the evaluation section we will compare our results with those of \cite{b48}. The authors in \cite{b48} use a low complexity method for matching D2D links with cellular UEs  to form partners for spectrum sharing. Another work we will compare with is \cite{b49}, which investigates the gain that cooperative multicast transmission provides when used to boost the data rate in D2D communication, enabling data sharing among users by implementing clusters.

All of the solutions discussed above solve only one of the many challenges identified in \ref{RC}, with the exception of \cite{b51}, which solves a joint sub-carrier assignment and power allocation problem.  There is also a yet unpublished work by \cite{b52}, which claims to be offering a solution in joint network admission control, mode assignment and power allocation in energy-harvesting D2D networks.

To the best of our knowledge there is currently no other work addressing 5G D2D communication issues using BDI agents with extended AI capabilities (BDIx).

\section{Distributed Control in D2D Through BDIx Agents}
In this Section we are describing both the new framework we are proposing for using BDI agents for D2D communication and we are also describing the DAIS Algorithm for selecting a node's transmission mode.

\begin{figure}[hbtp]
\centering
\includegraphics[width=0.9\linewidth]{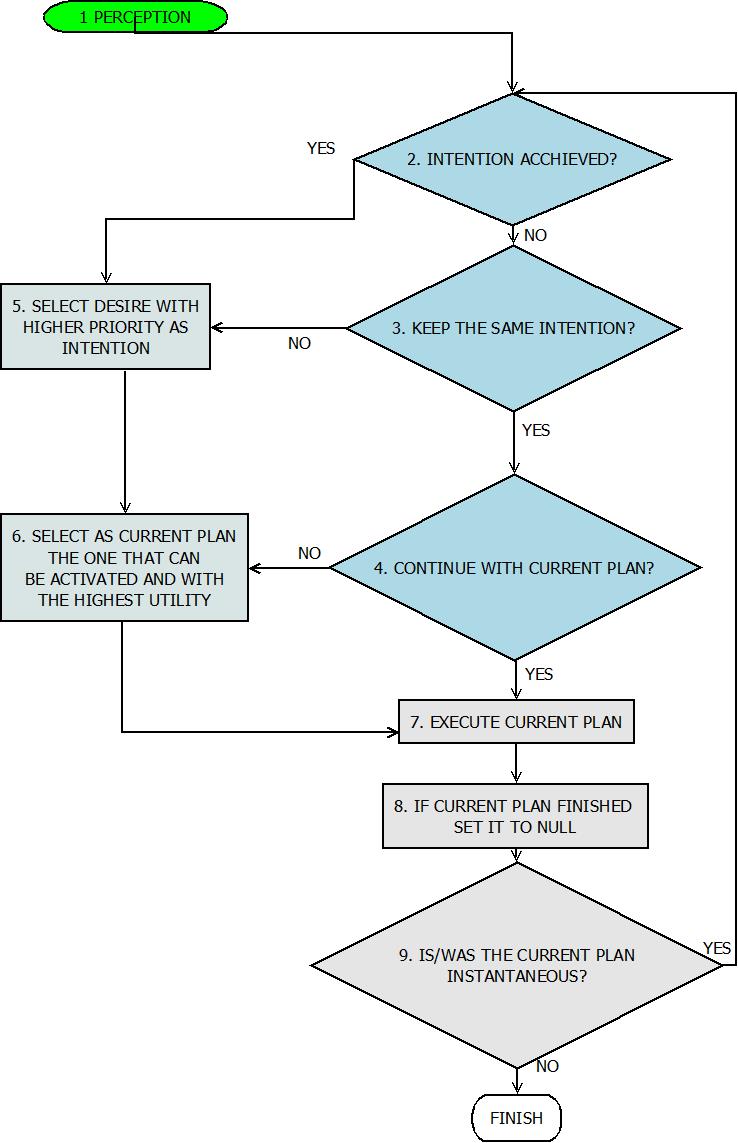}
\caption{Flowchart of BDIx Agent Operation}
\label{fig:flowchart}
\end{figure}

The flowchart in Figure \ref{fig:flowchart} shows the operation of a BDIx agent from the point it receives a message from the environment, until it selects and executes a plan.

After perceiving a change in its world, the agent checks if the Intention must be satisfied or must be changed. If the Intention is not changed then it continues with the execution of the Intention plan. Otherwise, the agent selects another Intention from the list that it has the higher priority and then it selects a Plan that will satisfy the selected Intention. After this it continues to execute the plan.

\subsection{Assumptions and Constraints}

The assumptions used in the design of the BDIx agents framework are the following:
\begin{itemize}
    \item The information needed by BDIx agents is the following: frequencies used, IP addresses, remaining energy, transmission mode (D2D Relay/D2D multi-hop/D2D cluster), etc (see Section \ref{TMode}).
	\item Location is known at the agent (all known devices have GPS).
	\item Location information and signals can be obtained within an operator's network.
	\item Each agent must be either a D2D Relay Node (D2D-R), a Multi-hop Relay Node (D2D-MHR), or a D2D Cluster Head (D2D-CH), or a "client" D2D node, i.e. at the edge of the communication path. So a D2D node can either serve or be served, not both. The agent will decide its role based on the beliefs and the events it has.
	\item A frequency should exist for the outband inter-communication between the BDIx agents.
	\item A threshold should be preset on Signal Quality (Received Signal Strength and Bit Error Rates)
	\item All D2D UEs that are in D2D-R or D2D-MHR transmission modes know their link and path rates and they can broadcast them over LTE proximity services.
	\item A BDIx Agent always accepts proposals from other BDIx Agents (e.g. a D2D UE to D2D-CH or D2D-MHR request is always granted).
	\item A BDIx Agent always selects unused RB (Resource Block in OFDMA). This is done for simplicity. The resource management and interference management will be done in future work.
    \item The UE device has two mobile interfaces or is using full duplex interface split equally between uplink and downlink.
	\item The UE device has one WiFi interfaces (like all mobiles).
\end{itemize}

\subsection{Sum Rate and Weighted Data Rate}\label{WDR}
One of the most common metrics for the evaluation of D2D solutions is Sum-Rate. The Sum-Rate is the total throughput in a network calculated as the sum of the data rates that are delivered to all UEs and D2D UEs in a network \cite{b53,b54}. Variations on Sum Rate exist, such as Weighted Sum-Rate in \cite{b55}, which considers certain links to be of more importance and gives different weights to the links based on the mode of transmission (direct, relay, etc).
We introduce a new metric called "Weighted Data Rate" (WDR). The WDR is defined at each node as the minimum data rate in the path that the UE selected. The minimum data rate of a path is the data rate of the weakest edge in the path. Our aim is, essentially, to maximize the WDR, i.e WDR = max(min(Link Rate) for each path.  The choice for using WDR instead of sum-rate is mainly for reducing the computational load of the BDI agent. The benefits will be shown clearly in the next section.

\subsection{The DAIS Algorithm for Transmission Mode Selection}

The following terms are used in the DAIS algorithm:
	\begin{itemize}

		\item \textbf{D2D-R} - D2D Relay node.
		\item \textbf{D2D-MHR} - D2D Multihop Relay node.
		\item \textbf{D2D-CH} - D2D Cluster Head.
		\item \textbf{WDR} - Weighted Data Rate.
		\item \textbf{MAXUsersCH} - Maximum Users Supported by CH = 255. This is based on WiFi Direct limits.
		\item \textbf{MAXQueryD2DRelayDistance} - Maximum distance to query D2D Relay UEs = 200m. It is the maximum distance of WiFi Direct (200m) or the maximum distance of LTE Direct (1000m).
		\item \textbf{MAXDistancetoFormCluster} - Maximum distance threshold to accept connection to the node if the UE is CH. This is the pre-defined maximum radius range between D2D UEs in order to form a D2D Cluster. This will be calculated based on the technology used (WiFi Direct or LTE Direct). It can be calculated from WiFi Direct range/2 (100m) or LTE Direct range/2 (500m).
		\item \textbf{MAXSpeedToFormBackhauling} - Maximum speed that a node is moving in order to be D2D Relay or D2D Multi Hop Relay = 1.5 m/s (pedestrian).
		\item \textbf{MAXDistanceMultiHop} - Maximum distance threshold for a UE from the nearest D2D Relay in order to act as D2D multi-hop relay. In order for the UE to select to be multi-hop, the device must have the weighted data rate to an existing D2D Relay greater than the weighted data rate of the existing D2D Relay.
		\item \textbf{MAXDistanceMoveAway} - Maximum Distance to move away from the current position in order to recalculate. It can be calculated from WiFi Direct range/2 (100m) or LTE Direct range/2 (500m).
		\item \textbf{PERCDataRate} - Percentage of difference of Data Rate in order to make D2D Relay connect from UE D2D multihop Relay to Gateway =20%
		\item \textbf{DeviceBatteryThreshold} - The minimum battery percentage in order of the D2D device to act as D2D-R or D2D-MHR is 70\%
        \item The D2D device power is calculated randomly and it is following a Gaussian distribution with mean of 0.6 and variance of 0.4.
	\end{itemize}

	\begin{table*}[h]
	\centering
	\caption{Algorithm Notations and Mathematical Representations}
	  \begin{adjustbox}{width=\textwidth}
	\begin{tabular}{@{} l | l }
		\toprule
		Notations  & Mathematical Representation     \\
		\midrule
		\textbf{d} & $\sqrt {\left( {UE{x_1} - D2D{x_2}} \right)^2 + \left( UE{{y_1} - D2D{y_2 }} \right)^2 }$ \\
		\textbf{maxD2DR} &   $\begin{multlined}[t][0.8\linewidth] D2D_{j}\  where\ WDR_{D2D_{j}}=(MAX (WDR_{D2D_{i}}) \exists\  D2D_{i}\  where\  d\geq\  MAXDistancetoFormCluster \\[-1ex] \wedge WDR_{D2D_{i}}\ \geq\ (WDR_{UE_{i}}\  +\ PERCDataRate*WDR_{UE_{i}}) \ \wedge \ i \in D2DR  \\
		\wedge \ COUNT ({D2D_{i}}_{g}\ \\[-1ex] WHERE\ g\ served by\ i) <= D) \\
		 \end{multlined}$  \\
		\textbf{maxD2DMHRNoConnections} &   $\begin{multlined}[t][0.8\linewidth] D2D_{j}\  where\ WDR_{D2D_{j}}=(MAX (WDR_{D2D_{i}})  \exists\  D2D_{i}\  where\  d\geq\  MAXDistancetoFormCluster  \wedge \\[-1ex] WDR_{D2D_{i}}\ \geq\ (WDR_{UE_{i}}\  +\ PERCDataRate*WDR_{UE_{i}}) \ \wedge \ i \in D2DMHR \ \wedge \ COUNT ({D2D_{i}}_{g}\ \\[-1ex] WHERE\ g\ served by\ i) = 0) \\ \end{multlined}$  \\	
		\textbf{maxD2DRNoConnectionsToBeD2DMHR} &   $\begin{multlined}[t][0.8\linewidth] D2D_{j}\  where\ WDR_{D2D_{j}}=(MAX (WDR_{D2D_{i}})  \exists\  D2D_{i}\  where\  d\geq\  MAXDistancetoFormCluster \wedge \\[-1ex] d\leq\  MAXQueryD2DRelayDistance   \   \wedge WDR_{D2D_{i}}\ \geq\ (WDR_{UE_{i}}\  +\ PERCDataRate*WDR_{UE_{i}}) \ \wedge \\[-1ex] \ i \in D2DR \ \wedge \ COUNT ({D2D_{i}}_{g}\ WHERE\ g\ served by\ i) = 0) \  \wedge \\[-1ex]  D2DDevicePower_{i} \geq\ DeviceBatteryThreshold \\  \end{multlined}$  \\						
		\textbf{maxD2DRToUseUED2DMHR} &   $\begin{multlined}[t][0.8\linewidth] D2D_{j}\  where\ WDR_{D2D_{j}}=(MAX (WDR_{D2D_{i}})  \exists\  D2D_{i}\  where\  d\geq\  MAXDistancetoFormCluster \ \wedge \\[-1ex]  d\leq\  MAXQueryD2DRelayDistance    \wedge WDR_{D2D_{i}}\ \ll\ (WDR_{UE_{i}}\  -\ PERCDataRate*WDR_{UE_{i}}) \ \\[-1ex]  \wedge \ i \in D2DR  \wedge D2DDevicePower_{i} \geq\ DeviceBatteryThreshold \\  \end{multlined}$ \\						
		\textbf{maxD2DMHRToUseAsMultiHop} &   $\begin{multlined}[t][0.8\linewidth] D2D_{j}\  where\ WDR_{D2D_{j}}=(MAX (WDR_{D2D_{i}})  \exists\  D2D_{i}\  where\  d\geq\  MAXQueryD2DRelayDistance \wedge \\[-1ex] d\leq\  MAXDistanceMultihop  \  \wedge WDR_{D2D_{i}}\ \geq\ (WDR_{UE_{i}}\  +\ PERCDataRate*WDR_{UE_{i}}) \ \wedge \\[-1ex] i \in D2DMHR \ \wedge  COUNT ({D2D_{i}}_{g}\   WHERE\ g\ served by\ i) = 0) \  \wedge \\[-1ex]  D2DDevicePower_{i} \geq\ DeviceBatteryThreshold \\  \end{multlined}$ \\							
		\bottomrule
	\end{tabular}
\end{adjustbox}
	\label{Table:MathRepresentation}
	\end{table*}

In our approach the D2D-R/D2D-MHR are using proximity services to broadcast the connection information (i.e. WDR, coordinates).

\begin{algorithm*}
                connect to BS (GateWay) \;

				\tcc{Check to find D2D Relay to connect as client}
				\tcc{Check if a D2D Relay device exists near the D2D UE with the maximum WDR}				
                \If{exists maxD2DR}
                { \tcc{Check to find D2D Relay to connect as client}
                                Connect UE as D2D Client to maxD2DR
                                using WiFi Direct\;
                }
				\tcc{Check if a D2D-MHR exists near the D2D UE with the maximum WDR and convert it to D2D-R }
                \ElseIf{exists maxD2DMHRNoConnections}
                { \tcc{Check to find D2D Multihop Relay that no one connects to, make it D2D Relay, and connect to it as D2D Client}
                                Request from maxD2DMHRNoConnections UE to be D2DR\;
                                Connect UE as D2D Client to maxD2DMHRNoConnections
                                using WiFi Direct\; \tcc{Now the D2D-MHR is D2D-R}
                }
                \tcc{Connect as D2D-R or Optimize a Path}
				\tcc{Check if a D2D-R device exists far from the D2D UE with maximum WDR and not have connections other than path to BS in order to connect to it as D2D-R (The device will convert to D2D-MHR)}
                \ElseIf{exists maxD2DRNoConnectionsToBeD2DMHR}
                { \tcc{Check to find D2D-R that no one connects to and make it D2D-MHR and connect to it as D2D-R}
                                Request from DMHRNoConnections UE to be D2D-MHR \;
                                Connect UE as D2D-R to
                                maxD2DRNoConnectionsToBeD2DMHR
                                using LTE Direct\; \tcc{Now the D2D-R is D2D-MHR and UE is D2D-R}
                }
                \tcc{Check if a D2D-R device exist far from the D2D UE with maximum WDR worse than the UE and with no connections other than a path to BS in order to connect to it as D2D-MHR (The device will connect as D2D-R to the new UE that is going to be D2D-MHR)}
                \ElseIf{exists maxD2DRToUseUED2DMHR}
                { \tcc{Check to find D2D-R that no one connects to, with worse WDR than the UE and make UE as D2D-MHR and ask the device D2D-R to connect to UE}
                                Set UE as D2D-MHR \;
                                Connect maxD2DRToUseUED2DMHR as D2D Relay to
                                UE using LTE Direct\; \tcc{Now the UE is D2D-MHR}
                }
				\tcc{Check if a D2DMHR device exist from the D2D UE with maximum WDR and no connections other than path to BS in order to connect to it as D2D-R}
                \ElseIf{exists maxD2DMHRToUseAsMultiHop}
                { \tcc{Check to find D2D-MHR that no one connects to, make UE as D2D-R and connect to it}
                                Set UE as D2D-R \;
                                UE.TransmissionMode=D2D Relay \;
                                Connect UE  as D2D-R to
                                maxD2DMHRToUseAsMultiHop using LTE Direct\; \tcc{Now the UE is D2D-R}
                }
                \Else
                {
                                set UE as D2D-MHR;
                                Stay connected to BS \;
                }

                \caption{DAIS Algorithm for Transmission Mode Selection Plan in BDIx Agents}
                \label{Algo1}

\end{algorithm*}

 The notation and mathematical representation of symbols used in the DAIS algorithm are presented in Table \ref{Table:MathRepresentation}. The plan of execution of transmission mode selection is shown in Algorithm \ref{Algo1}. This is executed at the startup phase of the BDIx Agent. The computational complexity of such an algorithm is O(n) because the algorithm calculates the values in Table \ref{Table:MathRepresentation} only once. In addition, the algorithm is quick because decisions are made locally and do not rely on global information. Since routes are created instantaneously and incrementally by each agent, by identifying local D2D-R and D2D-MHR using proximity services, the complexity is based on the actual number of D2D-R and D2D-MHR that the agent in each device must communicate with, whenever it is needed (e.g. in order to become D2D-R by connecting to an existing D2D-MHR in our algorithm).

\section{Performance Evaluation of the DAIS Algorithm}

In this section, we investigate the performance of the proposed DAIS algorithm.  The simulations are done using Java and Matlab.
We consider scenarios with one BS and a number of UEs ranging from 10 to 1000, over an area of 1000x1000 meters. The BS in the simulations is in the center of the grid. The simulation parameters are shown in Table \ref{Table:SimParams}. The parameters are taken from the standards for WiFi Direct \cite{b57}, LTE Direct \cite{b56}, and LTE communication \cite{b58,b59}.

\begin{table}[]
\centering
	\begin{tabular}{c|c}
		\hline
		\textbf{Simulation Parameters }              &\textbf{ Value}                  \\ \hline
		D2D power                       & 130 mW      \cite{b58,b59}              \\ \hline
		UE power                        & 260 mW      \cite{b58,b59}              \\ \hline
		WiFi Direct Radius              & 200 m       \cite{b56}                  \\ \hline
		LTE Direct Radius               & 1000 m      \cite{b57}                 \\ \hline
		BS Range                        & 1000 m      \cite{b58,b59}        \\ \hline
		Path loss exponent (Urban Area) & 3.5                    \\ \hline
		BS Antenna gain                 & 40 dB       \cite{b58,b59} 				\\ \hline
		UE/D2D antenna gain             & 2 dB        \cite{b58,b59} 				\\ \hline
		N0 (White Noise)	            & 0.0001			     \\ \hline
		D (WiFi Direct max clients)     & 200         \cite{b56} 	     \\ \hline
        N (no of UEs)                   & 10-1000               \\ \hline
		Shadowing					    & Log-normal   			 \\ \hline
		Mobility					    & Static scenario   	 \\ \hline		
	\end{tabular}
    \caption{Simulation Parameters}
	\label{Table:SimParams}
\end{table}

\begin{figure*}[h]
\centering
\begin{minipage}{.48\textwidth}
  \centering
  \includegraphics[width=\linewidth]{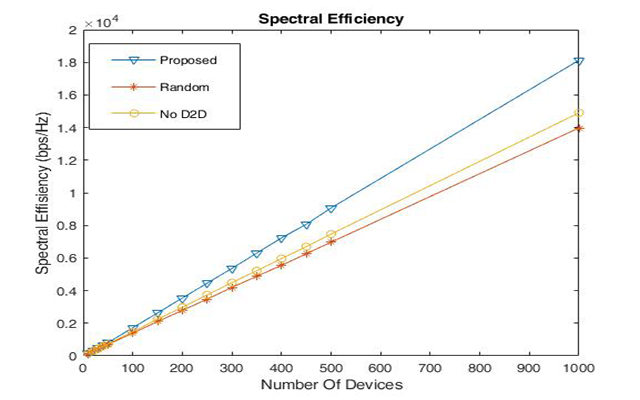}
  \captionof{figure}{Spectral Efficiency of Different Transmission Modes}
  \label{fig:2}
\end{minipage}%
\begin{minipage}{.48\textwidth}
  \centering
  \includegraphics[width=\linewidth]{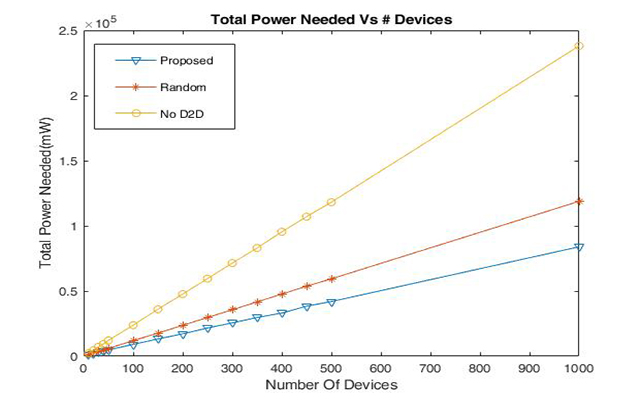}
  \captionof{figure}{Power Savings of Different Transmission Modes}
  \label{fig:1}
\end{minipage}
\end{figure*}

\begin{figure*}[h]
\centering
\begin{minipage}{.48\textwidth}
  \centering
  \includegraphics[width=\linewidth]{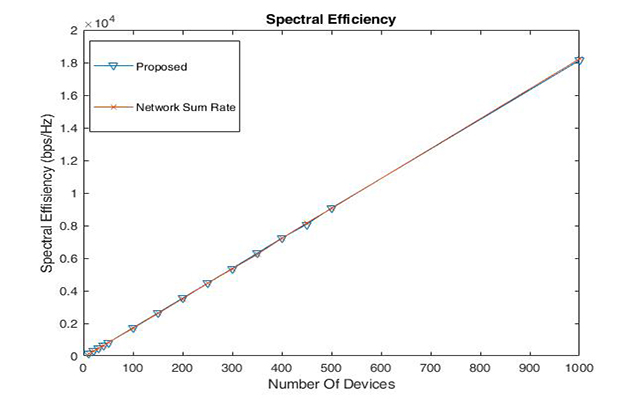}
  \captionof{figure}{Spectral Efficiency of Different Rate Options}
  \label{fig:4}
\end{minipage}%
\begin{minipage}{.48\textwidth}
  \centering
  \includegraphics[width=\linewidth]{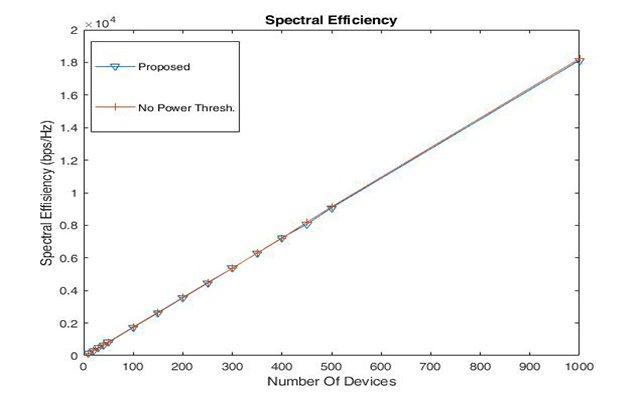}
  \captionof{figure}{Spectral Efficiency of Different Power Options}
  \label{fig:5}
\end{minipage}
\end{figure*}

The first thing we examine is the spectral efficiency of the proposed solution. Figure \ref{fig:2} shows that our proposed solution has a better performance compared to a random clustering solution and when no-D2D communication is used. The realized benefits are in the order of 30\%. The most interesting result is that random clustering results in spectral efficiency even worse than direct UE-BS communication.

Considering the power needed to realize the communication of the nodes, it is not surprising to see that clustering indeed requires less power. However, the proposed solution still outperforms the second best by about 25\%.

Within the proposed framework we have the ability to easily interchange metrics and parameters. In Section \ref{WDR} we have argued on the feasibility of using WDR instead of Sum-Rate in our calculations. Figure \ref{fig:4} shows that the use of WDR does not reduce the spectral efficiency of the system. The same happens if we consider an option in which a UE participates in the D2D communication depending on the remaining battery it has. Figure \ref{fig:5} shows no difference in spectral efficiency.

\begin{figure}[h]
	\centering
	\includegraphics[width=0.95\linewidth]{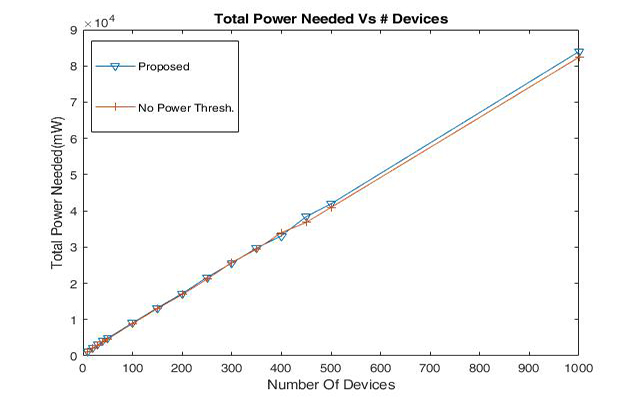}
	\caption{Power Saved}
	\label{fig:6}
\end{figure}

\begin{figure}[h]
	\centering
	\includegraphics[width=0.95\linewidth]{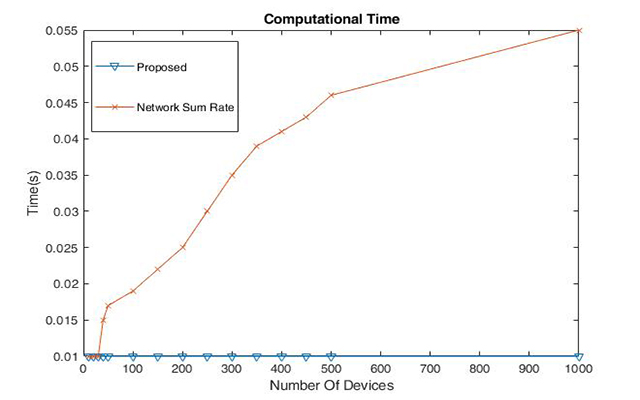}
	\caption{Computational Complexity}
	\label{fig:3}
\end{figure}

On the contrary, by utilizing a battery threshold we are slightly increasing the required power for the communication, as evident by the slight differences shown in Fig. \ref{fig:6}.

A significant result, which validates our choice of WDR is that the computational time needed to perform sum-rate calculations is up to five (5) times greater than the constant computation needed when we perform WDR calculations locally. This is ascribed to the fact that sum-rate needs to check all links in the network every time it needs to decide the transmission mode of a UE. As the number of UEs increases the computational time increases as well. In our case, the time to form a cluster is 100ms for any device density, because the D2D UEs have all their link rates precalculated, so that WDR for the new connection is easily computed.

By comparing the results of our approach with those in \cite{b49} we observe that for 50 UEs (maximum number considered in that work) we have the same number of clusters (seven) and the same amount of average UEs per cluster. However, we have no way of knowing if the solution in \cite{b49} can scale, whereas our approach is shown to scale well for at least up to 1000 UEs. In our approach the energy gained by the BS when we apply clustering is the same as in the work in comparison. However, in our case we can have 1000 UEs clustering at the almost instantaneous time of 100ms.
Another work that lends itself for comparison is \cite{b48}, when considered for similar BS and UE power as well as node density. The max number of UEs and D2D links used in that work is, again, in the order of 50.  In the best case scenario analyzed in \cite{b48}, the spectral efficiency reaches 220 b/s/Hz for N=30 UEs when all of them are D2D linked. The performance goes down to 180 b/s/Hz as the D2D links are reduced to twenty (20). By comparison, in our work, a similar number of UEs (N=30) and D2D links the corresponding spectral efficiency is 296 b/s/Hz. If we test it with 30 D2D links and 10 UE links the BDI solutions has a rate of 405 b/s/Hz which outperforms the 260 b/s/Hz of the paper in comparison.

\section{Conclusions and Future Work}
Device to Device (D2D) Communication is expected to be a core part of the forthcoming 5G Mobile Communication Networks. To achieve that goal, several challenges, like interference management, power control, and routing, among others, need to be addressed.  The paper investigates the problem of solving multiple D2D communication requirements in one framework by using BDI agents. Such agents can be implemented at the UEs and there is no need to change how BSs operate or to change the hardware at BSs or UEs. The current work focuses on the definition of a joint solution of D2D requirements. To that extend it contains a detailed proof-of-concept algorithm. which works towards deciding the transmission mode of each UE and forms the best possible paths towards the base station using relays and clusters. Through simulations the solution was found to ensure a high spectral efficiency and low computational load. In future work we will focus on the utilization of more AI approaches under our BDIx framework and we will evaluate a more dynamic environment by considering mobile UEs.

\balance

\end{document}